# Search for a light radion at HL-LHC and ILC250

F. Richard

Laboratoire de l'Accélérateur Linéaire (LAL), Centre Scientifique d'Orsay, Université Paris-Sud XI, BP 34, Bâtiment 200, F-91898 Orsay CEDEX, France

______________________________________________________________________

**Abstract:** *With the data collected by LHC at 13 TeV, the CMS collaboration has searched for low mass resonances decaying into two photons. This has resulted in the observation of 3 sd excess around 95 GeV, reminiscent of an indication obtained at LEP2 by combining the Higgs boson searches of the four LEP experiments. These observations, marginally significant, motivate the present study which shows how HL-LHC and ILC250 could search for a radion, the lightest new particle predicted within the Randall Sundrum (RS) model. ILC operating at a centre of mass energy of 250 GeV and with an integrated luminosity surpassing LEP2 by three orders of magnitude, could become the ideal machine to study a light radion and to observe, with very high accuracy, how it mixes with the Higgs boson and modifies the various couplings.*

## Introduction

After the discovery of the Higgs boson, the main issue remains to understand the hierarchy problem. So far there has been no firm experimental evidence indicating a possible solution to this puzzle. SUSY searches at LHC have covered a large, if not complete, mass domain without showing the beginning of an evidence in favour of this theoretically favoured solution. Theories with extra dimensions, less developed so far, offer another explanation. In particular, the Randal Sundrum (RS) model [1] uses an extra warped dimension to explains the huge hierarchy between the Planck mass and the electroweak scale. This model however is severely constrained by precision measurement and one needs either to assume very heavy, above 10 TeV, Kaluza Klein (KK) recurrences of ordinary bosons, therefore reintroducing a milder, but unexplained, mass hierarchy or to concoct sophisticated extensions of the initial scheme to reduce these lower limits. In these extensions it is conceivable but not certain that LHC could observe such recurrences. In addition to the KK recurrences, RS predicts the presence of a scalar particle, the so-called radion, which plays a fundamental role in stabilizing the compact extra dimension present in this theory. Noteworthy is the fact that such a particle could be the lightest extra particle predicted by this model. In the phenomenology attempting to explain the various anomalies observed in B physics [2], one would predict a radion ϕ lighter that 500 GeV and even below 100 GeV, so far not excluded by collider results.

In a recent paper [3], we have shown that such a light object could be cleanly observed in the process e+e-→Zϕ. This search is independent of the decay modes of ϕ and therefore offers the best prospects for light radions searches which are not necessarily detectable at LHC. The cross section for this process can be reduced with respect to the SM cross section for a Higgs boson, meaning that searches carried with LEP2 luminosities, with typically less than 1 fb-1 per experiment, are marginally relevant. Recall that Aleph had an evidence for a scalar at ~114 GeV, not confirmed by the other experiments. Putting together the four experiments [4], it was however possible to observe a modest indication, at the ~2.3 sd level, around 98 GeV.

At the future international linear e+e- collider ILC [5] operating at a centre of mass energy of 250 GeV and with an integrated luminosity surpassing LEP2 by three orders of magnitude, the precise measurement of the reaction e+e- -> ZH will allow to measure the HZZ coupling to ~0.5%, while the possibility of ϕ-H mixing could affect this coupling by a large amount. Combining this indirect method and the direct search, it has been shown in [3] that an almost full coverage of ϕ for masses below 150 GeV is provided by an e+e- operating at 250 GeV, the so-called ILC250.

LHC can search for this particle in various ways. Since it is expected that ϕ can couple more copiously to gluons than the SM Higgs, one expects a rather large production cross section. If the radion can decay into WW/ZZ or HH, there are good prospects of detection for ϕ heavier than 2MW.

For light radions one may use associated production with Z as proposed in [3] which predicts a substantial mass coverage. In some cases, the radion decays into b quarks, providing a similar search as the one used to detect the Higgs decay into bb. This method however has some limitations when, as is the case for a 98 GeV object, the mass falls very near to the Z mass and is therefore contaminated by the ZZ channel.

For light radions, as for the Higgs boson, a good detection prospect is through the two photons final state, since they provide a very precise mass reconstruction. Trigger limitations allow this type of search for masses above 70 GeV. Recently CMS [6] has released its results at 13 GeV and the data show an indication at ~3 sd around 95 GeV. Using the 'look elsewhere' criterion, this significance is clearly marginal. One may however wonder if this indication is consistent with LEP2, which would reinforce the case. The answer seems to be positive given that LEP2 mass resolution in four jets is clearly less precise than the resolution of CMS with two photons.

Awaiting for the ATLAS analysis, I will therefore investigate the consistency of the LEP2 and CMS indications in the framework of the RS model presented in [3]. In any occurrence, this analysis will remain valid for future searches of a light radion.

In the 1$^{st}$ section I will briefly recall the various RS models under consideration and the predictions for Higgs couplings. In the 2$^{nd}$ section, experimental observations gathered at LEP2 and LHC will be presented. I will discuss their compatibility in terms of the radion hypothesis. Future prospects will be presented in the 3$^{d}$ section in terms of LHC and ILC250.

It will be argued in section 4 that while an NMSSM model allows to reproduce the LEP2 behaviour with reduced coupling without altering the SM Higgs couplings, this model fails to explain that this extra scalar gives the same rate in the two photon channel as would the case for a SM Higgs boson.

# I The RS models

## I.1 Mass constraints

In this section, the various incarnations of the RS ideas are presented showing that the predicted masses of the new particles can vary over a wide range. In particular, the radion mass could be very light, of the same order as the Higgs mass, in which case one expects mixing effects which may result in major variations of the Higgs coupling, large enough to be measurable at LHC. It is however unlikely that H(125) is a pure radion state since its properties are too close to the SM predictions.

In addition to this effect, the corrections to the Higgs boson vacuum expectation value (VEV) induced by the KK gauge mixing can also be large as was first shown in reference [7]. This effect could also result in large corrections to the Higgs couplings if the KK gauge particles have masses below 10 TeV.

In the genuine version of RS, KK gauge particles give a large contribution to S,T parameters and need to exceed 10 TeV to pass the present constraints from LEP-SLD, therefore outside the reach of LHC but testable with e+e- precision accuracy on b and t couplings [8]. This state of affairs seems unsatisfactory given that it creates a new hierarchy between the masses of SM particles and RS excitations. For this reason, various extensions were invented to cure the problem.

With custodial symmetries, one can decrease the bound on KK gauge bosons to ~5 TeV at the limit of visibility of LHC [9]. Additional bosons and fermions without zero KK modes are then expected.

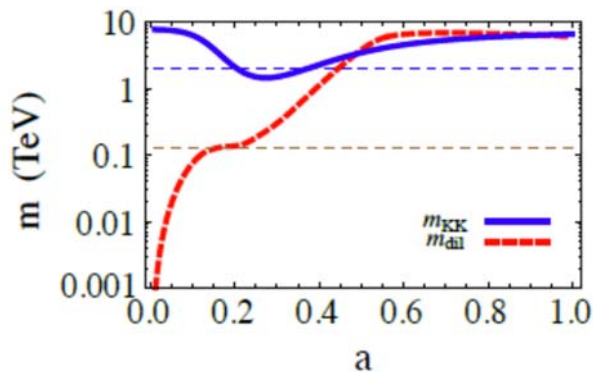

In the so-called soft-wall (SW) extension, one is able to modify the profile of the Higgs boson so as to reduce its overlap with KK particles which are peaked near the EW brane. Then the mass bound is reduced to 2 TeV for the KK gauge bosons and below 100 GeV for the radion, which makes such models particularly attractive to interpret the indications observed at LEP2 and CMS.

*Figure 1: The blue curve shows the minimum value for a KK vector boson in terms of an adjustable parameter a. The red dashed curve shows the predicted lower bound for the scalar particle.*

Figure 1 shows the mass bounds in terms of the parameter **a** which fixes the SW shape. Note that for a 95 GeV radion, **a**~0.1 , meaning that the minimal value for Mkk is ~5 TeV.

This scenario also claims to interpret B factory anomalies [10].

## I.2 Higgs couplings and RS effects

Higgs couplings can be influenced by the presence of KK vector bosons with masses below 10 TeV. This was first recognized in reference [7]. Figure 2 from [11] gives the ratio Rx=BRgg$\Gamma$x/SM of the $\gamma\gamma$ and ZZ* final states. These effects not only depend on Mkk but also on y*, an effective Yukawa coupling which drives the loop terms for $\Gamma\gamma\gamma$ and $\Gamma$gg.

For the coupling HZZ*, this effect gives a correcting factor **cz** which is model dependent. Assuming a custodial RS model, one has:

$$c_z = c_w = 1 - 0.078(5\text{TeV}/m_{g_1})^2$$

where $m_{g_1}$ stands for the mass of the first KK excitation of the gluon.

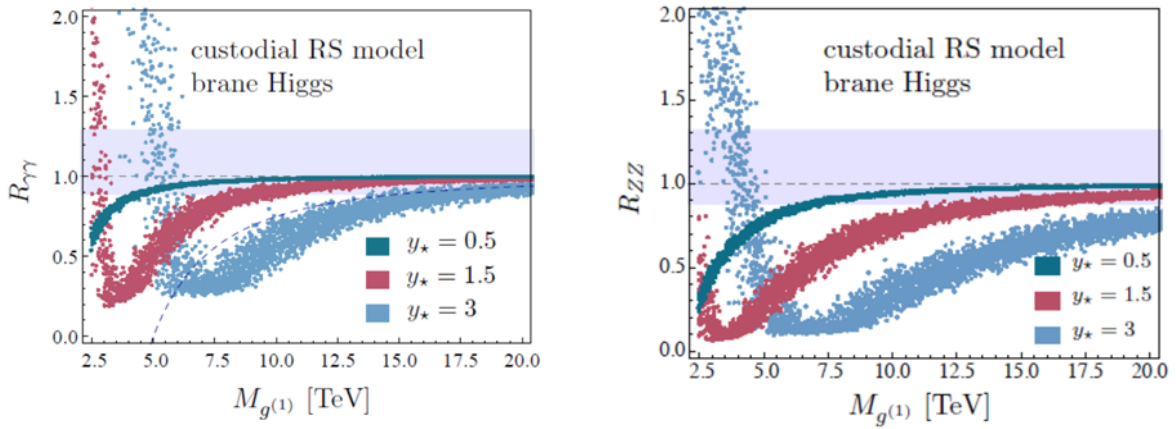

Figure 2: R$\gamma\gamma$ and Rzz variation with the KK gluon mass for 3 values of the Yukawa coupling y* defined in [11].

Heavy KK fermions will contribute to the loops of the gluon coupling. Contrary to the light fermions, where only the top quark contributes, they tend to have a universal localization in the warped space, near the EW brane, and therefore similar Yukawa couplings close to 1. The sum on all KK excitations gives a finite result which decouples for very large Mkk masses. Taking, tentatively, y*=1 for the Yukawa coupling, one gets the following correction factors [11]:

$$c_g = 1 - 0.33(5\text{TeV}/m_{g_1})^2 \quad c_\gamma = 1 + 0.13(5\text{TeV}/m_{g_1})^2$$

which can give larger deviations than for cz.

The b coupling, which dominates the total width, also depends on the Yukawa coupling y*. For y*=1 one has:

$$c_b = 1 - 0.05(5\text{TeV}/m_{g_1})^2$$

From the theoretical side, one needs to satisfy S,T constraints [11]. The allowed domain is shown in figure 3. This reference also gives $m_{g_1}$>4.8 TeV. These constraints drastically reduce the variations shown in figure 2. For instance, for $m_{g_1}$=5 GeV, figure 3 tells that y*<1.

From the present LHC data [12] shown in figure 4, one cannot exclude deviations with respect to the SM of the order 20-30%, which already gives a non-trivial limit on the parameters allowed to the radion and KK contributions. All other measurements on VH and WW fusion are still too imprecise to give a valid constraint.

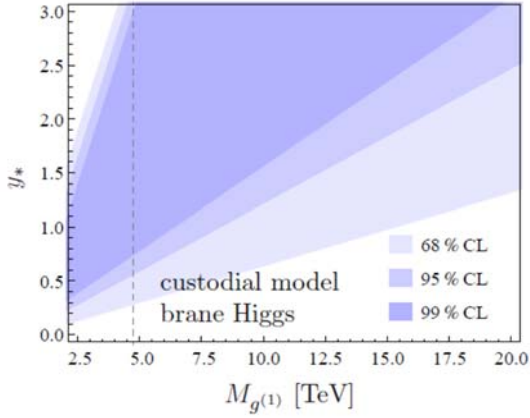 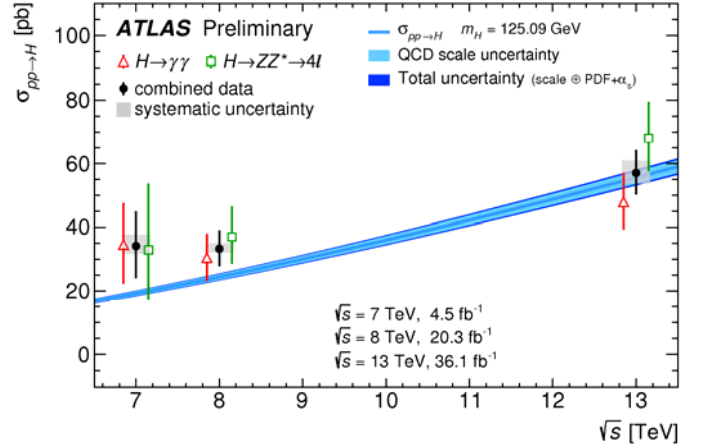

*Figure 3: Forbidden regions for $mg_1$ and $y^*$ in the custodial model of reference [11]*

*Figure 4: ATLAS measurements of the Higgs cross sections at 3 energies for the $ZZ^*$ and $\gamma\gamma$ final states.*

The Higgs-radion mixing calculations have been described in detail [3]. No attempt will be done to repeat these details sticking to the features needed to understand the phenomenology.

In figure 5 the blue curves assume $mg_1$ ~infinite. A large mixing solution, $|\xi|$~0.47, is needed to explain the observation of LEP2, as discussed in the next section. For this mixing parameter and taking Mkk infinite, one expects up to 10 to 20% deviation with respect to the SM. Given the present accuracies reached at LHC, both signs of $\xi$ are still acceptable although the positive sign seems to be disfavoured since it gives Rzz~0.8 while the 3 measurements shown in figure 4 indicate Rzz>1.

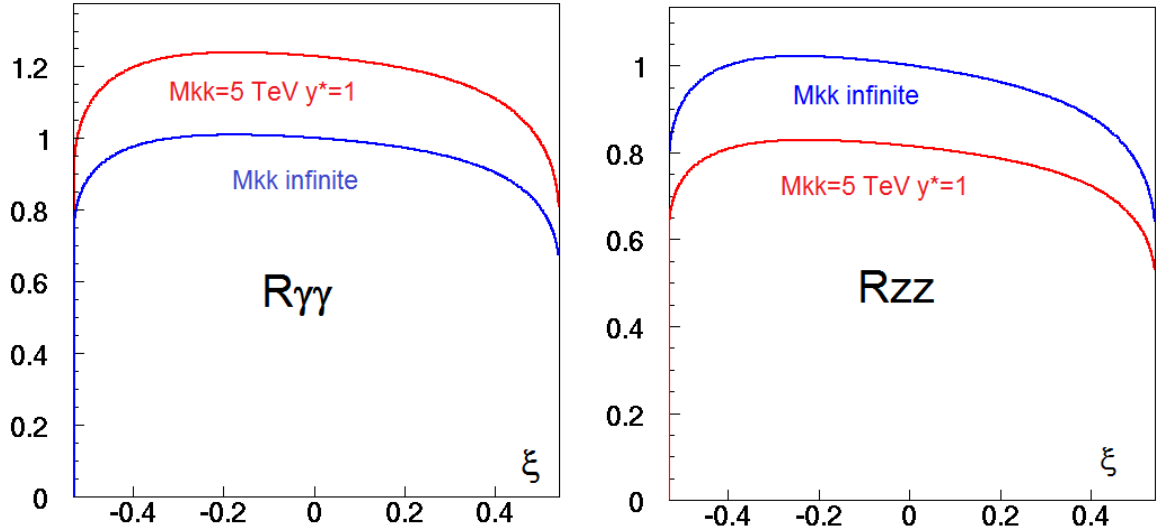

*Figure 5: Predicted values of $R\gamma\gamma$ and Rzz for the H(125) boson versus the mixing parameter $\xi$ for a vacuum expectation $\Lambda$=3 TeV. The red curves include the effects coming from contributions of KK vector bosons with 5 TeV mass, $y^*$ ~1 and $\Lambda$=3 TeV .*

In figure 5 the red curves are obtained with $y^*$~1 and $mg_1$~5 TeV and indicate maximal corrections. Again Rzz does not favour such a solution for both signs of the mixing parameters which implies that the KK vector bosons should be heavier than 5 TeV and/or $y^*$ well below 1.

The associated production ZH and Zϕ and the cross sections σ(pp->ZH) and σ(pp->Zϕ) are corrected by **cz²**. This is also true for σ(e+e- ->ZH) and σ(e+e- ->Zϕ).

If LHC discovers KK vector bosons, it will become possible to control **cz** but we will also need to know **y\*** to understand corrections on gluon and photon couplings. These quantities can also be determined from the Higgs couplings (section III.5).

# II The radion interpretation of LEP2 and CMS indications

LEP2 and CMS findings are recalled in figure 6. As expected in non-minimal Higgs models, one observes a reduced ϕZZ coupling with respect to Higgs SM coupling at the same mass but, surprisingly, the two photon mode exhibits the same coupling as for a genuine SM Higgs boson.

KK vector and germion contributions are also expected for the radion couplings, through their Higgs component. For the gluon and the photon couplings, there are large anomalous component which behaves like 1/Λ and should remain unaffected.

In section III.5, I will show how these various terms can be disentangled assuming a finite value of $mg_1$.

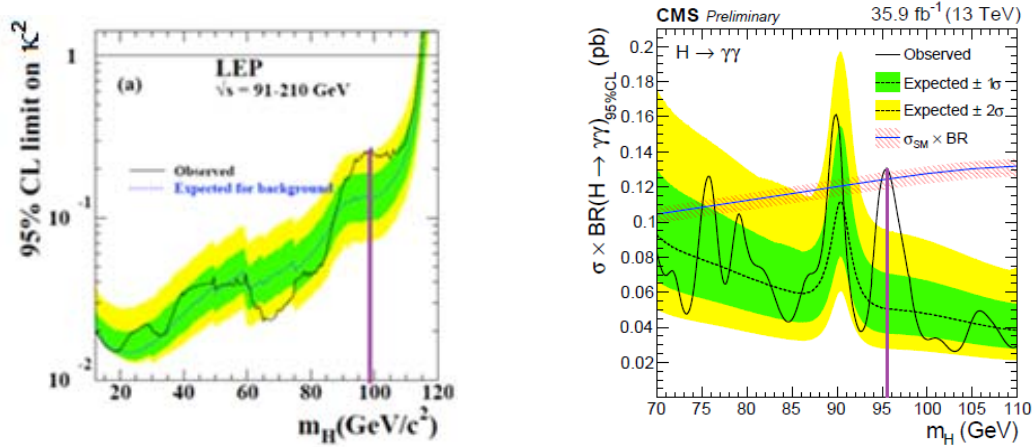

Figure 6: 95% CL limits from LEP2 and CMS on Higgs-like scalars versus mass. For CMS the hatched curve corresponds to a SM Higgs coupling. For LEP2 $\kappa^2$=1 corresponds to a SM Higgs coupling

For the time being, one assumes that KK vector bosons are heavy enough that one can neglect their contribution in e+e→Zϕ and gg→ϕ→γγ, such that the cross sections only depend on the vacuum term Λ and on the mixing parameter ξ.

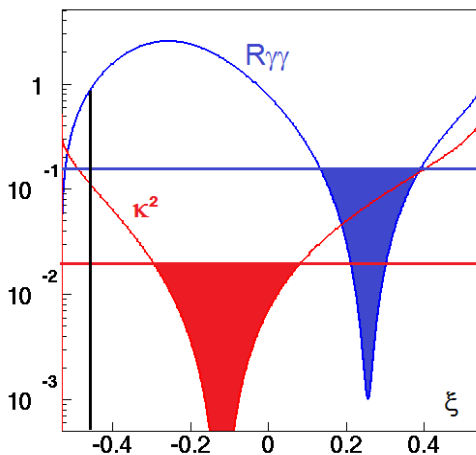

To ease the interpretation, cross sections will be normalized to the SM values. For LHC, one defines: Rϕγγ= BRϕggΓϕγγ/SM, where SM is the value expected for a hypothetical Higgs boson of 95 GeV with SM couplings. CMS data suggests that Rϕγγ ~1. LEP2 data suggest that $\kappa^2$~0.1-0.3 instead of 1 for a SM Higgs boson.
The two curves of figure 7, which are drawn with Λ=3 TeV, suggest that there are two solutions which corresponds to |ξ|~0.47. The two curves show two separated minima.

Figure 7: Rϕγγ and $\kappa^2$ predictions for ϕ(95) versus the mixing parameter ξ for a vacuum expectation Λ=3 TeV. Blind zones are in red for ILC and in blue for HL-LHC. ξ=-0.47 (black line) gives a solution consistent with LEP2 and CMS indications and with Higgs measurements.

For R$\phi\gamma\gamma$ the minimum comes from $\Gamma\phi\gamma\gamma$ which goes to zero, while the other minimum comes from the $\phi$ZZ coupling which goes to zero for a different value of $\xi$.

One can of course play with the parameter $\Lambda$. Reducing $\Lambda$ is marginally allowed by precision measurements as discussed in [3]. Increasing $\Lambda$ gives a worse agreement with the indications.

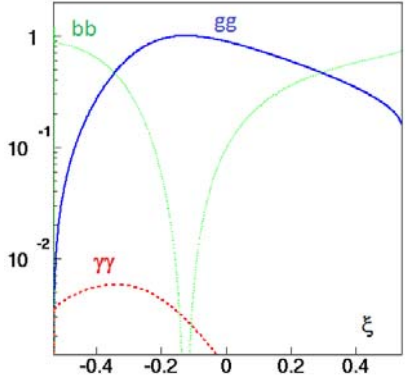

Figure 8: Branching ratios of $\phi(95)$ versus $\xi$ for $\Lambda$=3 TeV.

Figure 8 gives the branching ratio dependence on $\xi$. It shows that by measuring the branching ratio BR$\phi\gamma\gamma$ one can unambiguously distinguish between the two $|\xi|$~0.47 solutions.

Figure 7 illustrates the complementarity of HL-LHC and ILC250. For $\phi(95)$, ILC reaches its limit of sensitivity for $\kappa^2$~0.02 (horizontal red line), and HL-LHC for R$\gamma\gamma$~0.1 (horizontal blue line), then one sees that for the blind region of ILC (in red), LHC takes over. The same occurs for LHC (blue). This result assumes that LHC will collect 3000 fb-1 and that no systematic effect or trigger limitation will modify a naïve statistical extrapolation of present data.

For what concerns the WW fusion contribution, VBF, from LEP2 one predicts that this process is reduced by ~10 with respect to the SM process, meaning that the fusion cross section times the branching ratio into two photons should be ~0.15fb/10=0.015fb, therefore negligible with present luminosities. The CMS data show an indication which could be due to a wrong assignment of gg fusion events to VBF events. For what concerns VH, the cross section is further reduced by a factor 2, therefore negligible.

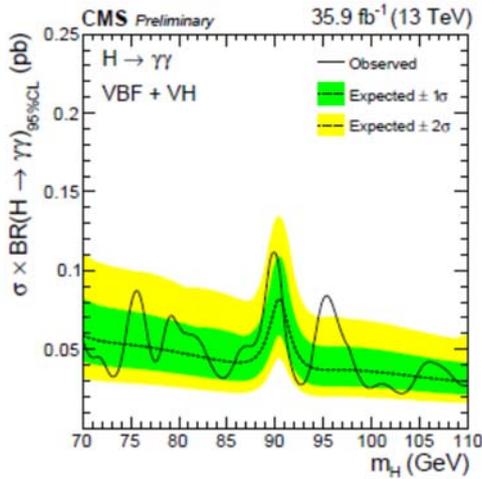

Figure 9: 95% CL for the VBF+VH processes.

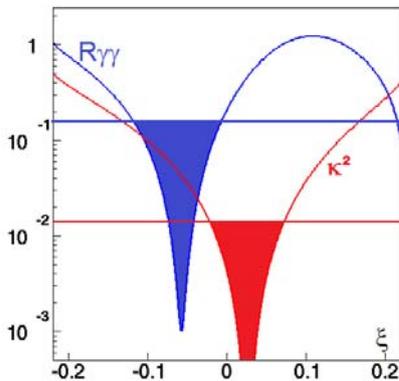

I have checked that for masses between 70 GeV and 150 GeV, ILC250 and HL-LHC provide complementary coverages for the direct search for a radion. Fig 10 shows R$\phi\gamma\gamma$ and $\kappa^2$ for $\phi(140)$ indicating that there is almost full coverage with HL-LHC and ILC250.

Figure 10: HL-LHC (blue) and ILC250 (red) coverage for radion searches for $\phi(140)$. Same conventions as for figure 7.

For masses below 70 GeV, one can try to use the process tt+$\phi$ using the decay into two photons. It turns out however that the $\phi$tt coupling behaves similarly to the $\phi$ZZ coupling and therefore is of no assistance in the dead region. Figure 10 from reference [3], shows that Higgs coupling measurements from ILC250 allow an almost complete coverage of this low mass region.

## III.1 Higgs coupling measurements

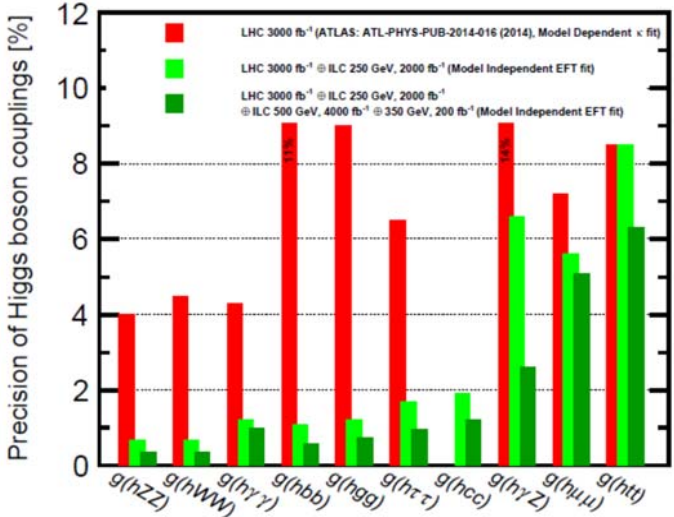

Figure11: Expected accuracies from ILC250 (light green) and ILC500 (dark green.

As shown in figure 11, ILC250 will provide very precise and complete measurements of the Higgs couplings [5], which will set constraints on the RS contributions. Assuming that KK bosons are very heavy, these measurements alone will allow extract the parameters $\xi$ and $\Lambda$. As an example, a measurement of the HZZ coupling, at the 0.6% level, is possible at ILC250 [5]. For the solution $\xi=-0.47$ the cross section should decrease by ~5%. Conversely one can deduce $\xi$ from this effect.

Could HL-LHC Higgs measurements be sensitive to such effects? Figure 5 suggests that deviations could, in the future, become significant.

## III.2 Radion measurements

The recoil mass method with e+e- $\rightarrow$ Z$\phi$, Z decaying into $\mu\mu$, allows to observe the mass distribution of figure 12. As for the ZH analysis described in [13], one uses the ZZ events to draw the expected background. As for the signal, one has to take into account the accuracy on the recoil mass, ~0.5 GeV, and the ISR and beamstrahlung effects [14]. Figure 12 says that, due to these radiation effects, about 50% of the $\mu\mu$ events remain in the peak region.

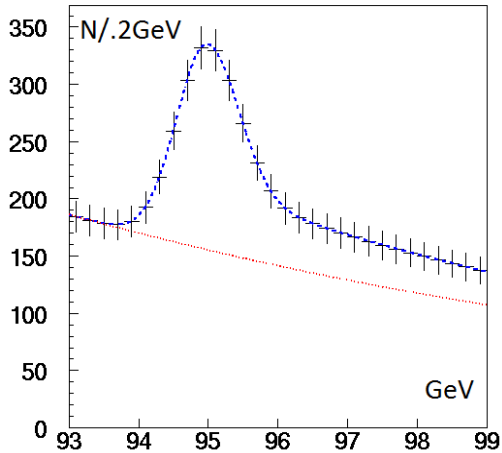

Figure 12: Predicted recoil mass distribution for the process e+e- ->Z$\phi$(95) using Z decays into $\mu\mu$. The dotted red curve is the background due to ZZ.

The measurements will be performed assuming that 2000 fb-1 can be collected at 250 GeV, according to the plans of ILC [5].

Expected accuracies are given in the following table. They have been extrapolated from the ILC analysis for the ZH channel.

The normalized coupling $\kappa$ and the mass be measured to:

$$\delta\kappa/\kappa = \pm 2\% \quad \delta M\phi = \pm 14 \text{ MeV}$$

assuming that the background is perfectly deduced from the ZZ events.

| Mode | bb | ττ | gg | γγ | Invisible |
|---|---|---|---|---|---|
| BR % | 78 | 10 | 11.5 | 0.5 | 0 |
| Relative Error % | 4 | 13 | 24 | 6 | <4 |

Again, these predictions ignore possible contributions from KK particles.

### III.3 $\Gamma\phi gg$ and $\Gamma\phi t$ from ILC and LHC measurements

At LHC, one is able to measure $BR\phi gg\Gamma\phi\gamma\gamma$ or, equivalently, $BR\phi\gamma\gamma\Gamma\phi gg$. As for the Higgs boson, the LHC accuracy is limited by the uncertainties on the gluon structure function ~3% and the theory uncertainties at ~5%. For the radion, it is reasonable to assume that these errors will be similar.

Using the ILC measurement of $BR\phi\gamma\gamma$ with a 6% error, one can deduce $\Gamma\phi gg$ with an accuracy of ~8%. This measurement would be the only one providing a result at the width level.

Knowing from ILC $BR\phi gg$ to 24% allows to extract the total width to a similar precision.

One can therefore expect to have:

$$\delta\Gamma\phi gg/\Gamma\phi gg \sim \pm 8\% \quad \text{and} \quad \delta\Gamma\phi t/\Gamma\phi t \sim \pm 24\%$$

Other LHC measurements are still very limited in accuracy. For instance, the process $\phi(95)tt$ would be about four time smaller than Htt which itself has still very large errors. It is not excluded that at HL-LHC, this situation could improve, providing valuable new inputs.

### III.4 $\xi$ and $\Lambda$ with Mkk=∞

For the radion, the most precisely measured quantities are $\kappa$, $BR\phi bb$, $BR\phi\gamma\gamma$ and $\Gamma\phi gg$. One can as well use the Higgs couplings which are far more precisely known.

To get a feeling on the size of the accuracies on the radion parameters for Mkk infinite, one can select two observables $\Gamma\phi gg$ and the coupling of the radion to ZZ, known to respectively 8% and 2%.

To satisfy the observed LHC and LEP2 indications, one takes $\Lambda$=3 TeV and $\xi$=-0.475, which gives a mixing angle $\theta$=0.67 rad. One then predicts the following accuracies:

$$\delta\xi = \pm 0.01 \quad \delta\Lambda/\Lambda = \pm 2.4\%$$

Expected deviations from the Higgs couplings, together with the expected errors at ILC250, are given in the next table.

| H coupling | ZZ | bb | ττ | gg | γγ |
|---|---|---|---|---|---|
| Deviation % | 2.2 | 7 | 5 | -1.1 | 0.3 |
| Relative Error % | 0.6 | 1 | 1.7 | 1 | 1 |

The main effect is seen on fermion couplings which will influence the total width and therefore $R\gamma\gamma$ and $R_{ZZ}$. It is however fair to say that this effect depends on the location of bL and bR in the extra dimension, which is still unknown. This location depends on the parameters cL and cR and one needs to know cL-cR. I have taken this difference equal to 1 for b quarks and 0.5 for $\tau$ leptons.

### III.5 $\xi$ and $\Lambda$ for $mg_1$=5 TeV and $y^*$=1

For mg1=5 TeV and y*=1, Higgs couplings would be significantly altered:

| H coupling | ZZ | Bb | $\tau\tau$ | Gg | $\gamma\gamma$ |
|---|---|---|---|---|---|
| Radion+KK | -6 | -2 | 0 | -34 | 15 |
| Relative Error % | 0.6 | 1 | 1.7 | 1 | 1 |

These large effects mean that Higgs precision measurements allow to set a stringent bound on $mg_1$:

**$mg_1$>17.5 TeV at 90% CL**

From the ILC+LHC data one can extract the various parameters of the model with the accuracies reported in the following table.

| $mg_1$ TeV | y* | $\delta\Lambda/\Lambda$ TeV | $\xi$ |
|---|---|---|---|
| 5±0.2 | 1±0.04 | 3% | -0.47 ± 0.02 |

$mg_1$ and y* determinations rely on [11] and are therefore model dependent.

## IV An NMSSM interpretation

There are alternate explanations for the presence of two scalars H2(125) and H1(95), as for instance

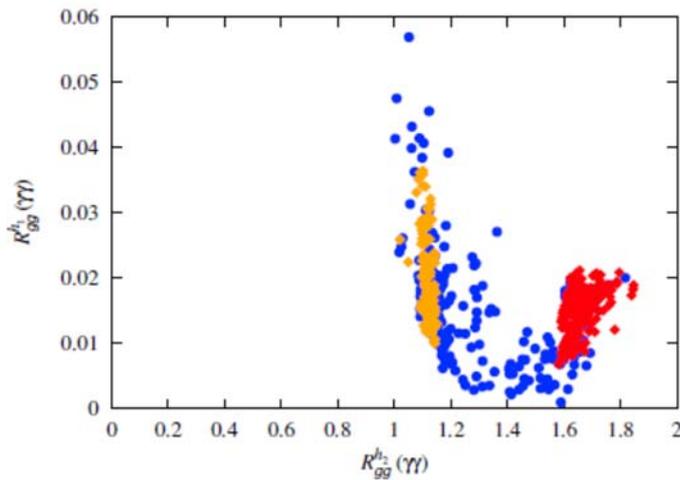

in the NMSSM model proposed in [15]. Can one distinguish between the two interpretations?

While [15] can reproduce the LEP2 observation, for LHC it predicts $R\gamma\gamma$ much lower than 1, contrary to observation.

This model predicts several other scalar particles, $H_3$, $a_1$, $a_2$ and $H^\pm$ with masses below 1 TeV, which, so far, have not been seen.

*Figure 13: $R\gamma\gamma$ NMMSM predictions for H2(125) and H1(95).*

# Summary and conclusion

Reported indications from LEP2 and CMS are compatible with a light radion predicted within RS. This radion should be strongly mixed with the Higgs boson. This implies that there could be corrections to the most precisely measured processes pp->H->$\gamma\gamma$ and pp->H->ZZ*.

If confirmed, this radion would be ideally measured in the reaction e+e-→Z$\phi$ at ILC250, giving the $\phi$ZZ coupling, the mass, the various branching ratios, including the invisible one. In combination with LHC, one would be able to determine $\Gamma\phi$gg and the total width.

From the Higgs and radion measurements, one can precisely deduce the two parameters of the model $\xi$ and $\Lambda$ and the mass of the first KK boson excitation or set a lower limit on its value, close to 20 TeV.

Generally speaking, ILC250 together with HL-LHC allows to search for a light radion between 70 GeV and 150 GeV. Below 70 GeV one can use the Higgs measurements at ILC250 to complement the direct searches.

An interpretation within NMSSM of the CMS indication seems excluded by the large two photon cross section observed by CMS.

Data collected at 13 TeV by ATLAS and CMS during the period 2016-2017 correspond to four times the sample analysed by CMS and should provide the last word on the experimental indication from CMS.

In any occurrence, the present analysis is meant to serve as an input for future searches, direct or indirect, for a light radion at HL-LHC and ILC250. This is just the beginning of a great challenge for particle physics.

**Acknowledgements:** It is a pleasure to thank Abdelhak Djouadi and Gregory Moreau for their encouraging comments. Special thanks to Andrei Angelescu for sharing his precious expertise on this topic.